\documentclass[journal]{IEEEtran}

\ifCLASSOPTIONcompsoc
  \usepackage[nocompress]{cite}
\else
  \usepackage{cite}
\fi

\ifCLASSINFOpdf
  \usepackage[pdftex]{graphicx}
\else
  \usepackage[dvips]{graphicx}
\fi

\usepackage{algpseudocode}
\usepackage{xspace}

\usepackage{multirow}

\usepackage{cases}
\usepackage{array}
\usepackage{color}
\definecolor{mycolor}{rgb}{1,0,1} %{.5,0,0} %
\usepackage{url}
\newcommand \ignore[1]{}
\algnewcommand\And{\textbf{and}\xspace}

\begin{document}
\title{Summary: Multi-modal Biometric-based Implicit Authentication of Wearable Device Users}
%
% Reason for Multi-modal >>> biometrics are from different sources, e.g., HR vs Steps; Generic >>> instead of creating models for different non-sedentary activity levels ONLY one sedentary model with levels as a feature, also, the same feature set and model across all subjects, also, relatively simple model with low feature count; Implicit >>> users' don't need to perform any explicit actions like tying/writing/motion

\author{Sudip~Vhaduri and~Christian~Poellabauer,~\IEEEmembership{Senior~Member,~IEEE}% <-this % stops a space
\IEEEcompsocitemizethanks{\IEEEcompsocthanksitem The authors are with the Department of Computer and Information Science, Fordham University, NY, USA and Department of Computer Science and Engineering, University of Notre Dame, IN, USA.\protect\\
%E-mail: \{svhaduri,cpoellab\}@nd.edu
E-mail: vhaduri@gmail.com, cpoellab@nd.edu
}
%\thanks{Manuscript received April 19, 2016; revised August 26, 2016.}
}

\ifCLASSOPTIONpeerreview
    \markboth{Journal of \LaTeX\ Class Files,~Vol.~14, No.~8, August~2015}%
    {Shell \MakeLowercase{\textit{et al.}}: Bare Demo of IEEEtran.cls for IEEE Journals}
\fi

\maketitle

% \fi
\IEEEpeerreviewmaketitle

\section{Introduction}\label{introduction}

With the emergence of the Internet of Things (IoT), we now have access to a multitude of devices with advanced 
capabilities that allow us to remotely collect information or control physical objects. Examples of such systems include alarm systems, entertainment devices, vehicles, smart buildings, to name a few. 
At the same time, smartphones and wearables have also advanced in their sensing and computational capabilities, which enable many new applications and usage scenarios. While smartphones are already widely used, wearables are still growing in popularity with the arrival of new applications, including the ability to identify a user
to third party services, store sensitive user information (i.e., passwords, credit card information), unlock vehicles, access phones and other paired devices, manage financial payments, monitor or track individuals (e.g., child monitoring or fall detection), and assess an individual's health and fitness.

However, wearables also raise new challenges, specifically in terms of security. Unauthorized access of a wearable can enable access to other sensitive IoT objects, which poses a significant risk~\cite{zeng2017wearia}. 
Unauthorized users could also access data on the wearables, e.g., many applications and services provided by a wearable depend on sensor and user data stored on the device.
Another concern is the reliability (i.e., trustworthiness) of the physiological and activity data collected by wearables, e.g., many healthcare providers and researchers rely on wearables to monitor their patients or study participants remotely, where users may be tempted to give their own devices to others, e.g., to reach a prescribed amount of activity or to contribute the required amount of data to maintain compliance and receive financial incentives. Therefore, there is a need for robust and accurate authentication mechanisms specifically for wearable device users.

Existing wearable device authentication mechanisms are often knowledge-based regular PIN locks or pattern locks~\cite{nguyen2017smartwatches}, which suffer from a scalability issue~\cite{unar2014review}, i.e., with an increasing reliance on protected devices, a user is often flooded with passwords or PIN requests to obtain access to various data and services. 
Knowledge-based approaches also require users to interact with the display, which may either be inconvenient or even completely absent in many wearables~\cite{unar2014review,zeng2017wearia}. Many times, users opt to completely disable security mechanisms out of convenience.
Furthermore, knowledge-based approaches suffer from observation attacks such as shoulder surfing~\cite{unar2014review}.
Therefore, in recent years, biometric-based solutions have been proposed, since they provide opportunities for {\em implicit authentication} by removing direct user involvement or attention~\cite{unar2014review,zeng2017wearia}.
A market intelligence firm has also predicted that annual biometric hardware and software revenue will grow at a compound annual growth rate (CAGR) of 22.9\% from \$2.4 billion in 2016 to \$15.1 billion worldwide by 2025~\cite{biometric_revenue}, which further provides evidence that there is an opportunity to utilize biometric-based authentication.

However, biometric-based authentication also has challenges and shortcomings, specifically in terms of accuracy and usability. For example, behavioral biometric-based approaches such as gait or gesture analysis often fail to authenticate a user during sedentary periods~\cite{zeng2017wearia}.
%\textcolor{mycolor}{
Due to the limited computational capabilities and energy resources, most traditional user authentication approaches, e.g., using fingerprints, voice, breathing patterns, keystroke dynamics, head or arm movements, gait, electroencephalography (EEG), and electrocardiograms (ECG) are often not suitable for wearable devices. Furthermore, low-cost sensors in wearables may be less accurate (leading to noisy data recordings) or collect recordings very infrequently (e.g., only once per minute).
The limited display sizes of wearables add another constraint that limits the choices of authentication mechanisms. 
%}

The main contribution of this paper is an implicit wearable device user authentication approach using three different types of biometric data: behavioral biometrics (step counts), physiological biometrics (heart rate), and hybrid biometrics (calorie burn and metabolic equivalent of a task or MET), all of which are easily obtainable in many state-of-the-art wearables. Our approach is to authenticate a user based on coarse-grained (i.e., one sample per minute instead of multiple samples per second or millisecond) processed (i.e., not raw) biometric data. 
We train and test different authentication models with different feature sets using a Support Vector Machine (SVM) classifier, which was found to be the most accurate in our previous work~\cite{vhaduri2017wearable}. 
Our analysis using data from over 400 Fitbit users shows that our multi-biometric-based implicit approach is able to authenticate subjects with an average accuracy of about .93 (sedentary) and .90 (non-sedentary). 
We also find that the hybrid biometrics (i.e., calorie burn and MET) perform the best, whereas the behavioral biometrics (i.e., step counts) do not have a significant impact on authentication even during non-sedentary periods. Using error analysis we demonstrate the trade-off between usability and security of our authentication approach. 
\section{Approach}

\begin{table*}[!t]
\caption{Candidate feature set consisting of 27 features computed from each of the four biometrics}
\label{table:init_featureset}
\centering
\begin{tabular}{l}
\hline
mean ($\mu$), standard deviation ($\sigma$), variance ($\sigma^2$), coefficient of variation ($cov$), maximum ($max$), minimum ($min$), range ($ran$), coefficient of \\
range ($coran$), percentiles ($25^{th}$ ($p25$), $50^{th}$ ($p50$), $75^{th}$ ($p75$), and $95^{th}$ ($p95$)), inter quartile range ($iqr$), coefficient of inter quartile range \\
($coi$), mean absolute deviation ($mad\_\mu$), median absolute deviation ($mad\_Mdn$), mean frequency ($f\_\mu$), median frequency ($f\_Mdn$), \\ 
power ($P$), number of peaks ($np$), energy ($E$), root mean square ($rms$), peak magnitude to rms ratio ($p2rms$), root sum of squares ($rss$), \\
signal to noise ratio ($snr$), skewness ($\gamma$), kurtosis ($\kappa$) \\
\hline
\end{tabular}
\end{table*}

%\subsection{NetHealth Study Dataset}\label{study}
\noindent{\bf NetHealth Study Dataset:}
The {\em NetHealth} mobile crowd sensing (MCS) study~\cite{vhaduri2016assessing,vhaduri2016cooperative,vhaduri2017discovering,vhaduri2017wearable,vhaduri2017towards,vhaduri2018hierarchical,vhaduri2018biometric,vhaduri2018impact,vhaduri2018opportunisticICHI,vhaduri2018opportunisticTBD,vhaduri2016design,vhaduri2016human,vhaduri2017design,vhaduri2019multi} began at the University of Notre Dame with over 400 freshmen (age 18 $\pm$ 1 years) recruited from the 2015 class.
Subjects were instructed to continuously wear a Fitbit Charge HR device, which provided minute-level heart rate, calorie burn, metabolic equivalent of task or MET, physical activity level/intensity (e.g., sedentary, light, fair, and high), step count, sleep status, and self-recorded activity labels. 
These collected data can be divided into three biometric groups: behavioral (e.g., step counts, activity level/intensity), physiological (e.g., heart rate) and hybrid (e.g., calorie burn, MET) biometrics.

\noindent{\bf Pre-Processing and Feature Computation:}
We first remove periods of activity minutes that do not match with periods of heart rate measurements. A similar filtering approach is applied for calorie burn and MET.
Next, we segment continuous biometrics into five-minute non-overlapping windows starting from a change of activity levels. Each window contains five consecutive samples recorded at the same activity level. 
Table~\ref{table:init_featureset} presents the candidate feature set.
We compute 108 features for each window, when considering all four biometrics together. For {\em non-sedentary} periods, we also consider the activity level (i.e., light, fair, and high) as an additional feature.
Each biometric is referred to by its initial: ``C'' (calorie burn), ``S'' (step counts), ``M'' (MET), and ``H'' (heart rate). Combinations of these letters are used to represent the corresponding combinations of the biometrics, e.g., ``CH'' represents a combination of calorie burn and heart rate.

%\subsection{Feature Selection}\label{feature_selection}
\noindent{\bf Feature Selection:}
First, we apply the {\em Kolmogorov-Smirnov} (KS)-test to select significant features from the candidate feature sets consisting of 27 features from each of the four biometrics. 
Then, we apply two separate approaches to reduce the feature count -- (1) remove redundant features using the {\em Pearson Correlation} (PC)-based approach and (2) reduce the feature count using {\em Standard Deviation} (SD)-based feature selection approach.

\noindent{\bf Authentication Models:}
We use: (1) the binary {\em Quadratic Support Vector Machine} (q-svm), i.e., SVM with a second order polynomial kernel function defined as $K(x_i,x_j) = (1 + \gamma x_i^Tx_j)^d$ and (2) the unary {\em Gaussian Support Vector Machine} (g-svm), i.e., SVM with the Gaussian or Radial Basis Function (RBF) defined as $K(x_i,x_j) = exp(-\gamma x_i^Tx_j)$, where kernel scale parameter $\gamma = 1$, degree $d = 2$, and $x_i$ and $x_j$ are two feature vectors/windows. We also set the miss-classification cost/penalty, $C = 1$. We perform wearable device user authentication separately for sedentary and non-sedentary periods.

%\subsection{Analysis Using Authentication Models}
\noindent{\bf Train-Test Sets:}
For each feature set with $N$ subjects, we build $N$ separate models, one for each subject.
When using the binary q-svm classifier, we train a model with 75\% data of one subject (positive class) and 75\% data from the rest of the $N-1$ subjects (negative class). However, when using the unary (i.e., 1-class) g-svm classifier, we train a model using a subject's own data with a certain percentage of data being considered as outliers. For both types of classifiers, we test a model on of the $25\%$ data both from positive and negative classes. In all cases, we use balanced datasets. 

%\subsection{Performance Measures}
\noindent{\bf Performance Measures:}
In addition to standard {\em Accuracy (ACC)}, {\em False Positive Rate (FPR)}, and {\em False Negative Rate (FNR)}, we also use the {\em Equal Error Rate (EER)}, which is defined as the point when FNR and FPR are equal, i.e., a trade-off between the two error measures (i.e., FNR and FPR). 
Note that literature often also uses {\em False Acceptance Rate (FAR)} and
{\em False Rejection Rate (FRR)}, which are exactly the same as FPR and FNR, respectively.
\section{Results}

\subsection{Comparing Biometrics and Feature Selection Approaches}\label{best_bio_per_selec_app}

Table~\ref{table:acc_summ_feat_sets} summarizes the findings when applying q-svm on the three feature selection approaches. In the table, for each feature selection approach, we present the best biometrics, i.e., biometric combinations that optimize feature count and classification accuracy. Also, $n$, $N$, and $W$ stands for the number of features in a feature set, the number of subjects used for training-testing, and set of random windows picked from a subject, respectively. In the table, we observe that the KS- and SD-based approaches achieve similar performance. 

In Table~\ref{table:acc_summ_feat_sets} we observe that in general hybrid biometrics (i.e., ``C'' and ``M'') perform better than other behavioral (``S'') and physiological (``H'') biometrics, which happens because hybrid biometrics are derived from both behavioral and physiological biometrics in addition to demographic data such as weight and age of a user. 
%Therefore, it can be a better idea to develop authentication systems based on hybrid biometrics such as calorie burn and MET instead of existing behavioral or physiological biometrics-based approaches.

\begin{table}[b]
\caption{Authentication Summary}
\label{table:acc_summ_feat_sets}
\begin{center}
\begin{tabular}{l|c c c|c c c}
\hline
%\cline{1-7}
   &  \multicolumn{5}{c}{Feature selection approach}  \\
\cline{2-7}
   &  \multicolumn{3}{c|}{{\em Sedentary}}  & \multicolumn{3}{c}{{\em Non-sedentary}}  \\
\cline{2-4}\cline{5-7}
                & KS    &  PC   &  SD   & KS    & PC &  SD \\
\hline
%Best biometrics & CM    &  CMH  &  CM   & CM    & CMH & CMSH \\
Best            &       &       &       &       &     &  \\
biometrics      & CM    &  CMH  &  CM   & CM    & CMH & CSMH \\
%\hline
%                &       &       &  CSM, &       &     &  \\
%Similar         &       &       &  CMH, &       &     &  \\
%biometrics      & CSM   &  CSMH &  CSMH & CMH   & CM  & CMH \\
\hline
%-------------------including FREQ----------------------------
%$n$             & 53    &  57   &  22   & 30    & 22  & 33 \\
%-------------------excluding FREQ----------------------------
$n$             & 49    &  53   &  20   & 27    & 21  & 30 \\
%-------------------Incorrect N---------------------------------
%$N$             & 412   &  411  &  371  & 332   & 332 & 373 \\
%-------------------Correct N-----------------------------------
$N$             & 412   &  411  &  415  & 413   & 413 & 412 \\
$|W|$           & 544   &  637  &  370  & 331   & 331 & 372 \\
\hline
&  \multicolumn{6}{c}{Binary Classification}  \\
\hline
%-------------------including FREQ----------------------------
%-------------------excluding FREQ----------------------------
$\mu(ACC)$      & .93  & .86   &   .93  & .90  & .86  & .88 \\
$\sigma(ACC)$   & .03  & .04   &   .04  & .04  & .05  & .05 \\
\hline
$\mu(FNR)$      & .04  & .15   &   .03  & .06  & .10  & .09 \\
$\sigma(FNR)$   & .03  & .05   &   .03  & .04  & .05  & .04 \\
\hline
$\mu(FPR)$      & .10  & .14   &   .11  & .14  & .18  & .15 \\
$\sigma(FPR)$   & .05  & .05   &   .06  & .05  & .06  & .07 \\
\hline
&  \multicolumn{6}{c}{Unary Classification}  \\
\hline
%-----------------------------------------------------------------
$\mu(ACC)$      & .88  &  .76  &   .90  & .74  & .64  & .68 \\
$\sigma(ACC)$   & .07  &  .06  &   .07  & .06  & .07  & .07 \\
\hline
$\mu(FNR)$      & .14  &  .27  &   .09  & .16  & .07  & .07 \\
$\sigma(FNR)$   & .08  &  .11  &   .07  & .07  & .04  & .03 \\
\hline
$\mu(FPR)$      & .11  &  .21  &   .12  & .36  & .64  & .56 \\
$\sigma(FPR)$   & .16  &  .16  &   .16  & .13  & .14  & .14 \\
\hline
\end{tabular}
\end{center}
\end{table}

\begin{figure}
\centering
  \includegraphics[width=.95\linewidth]{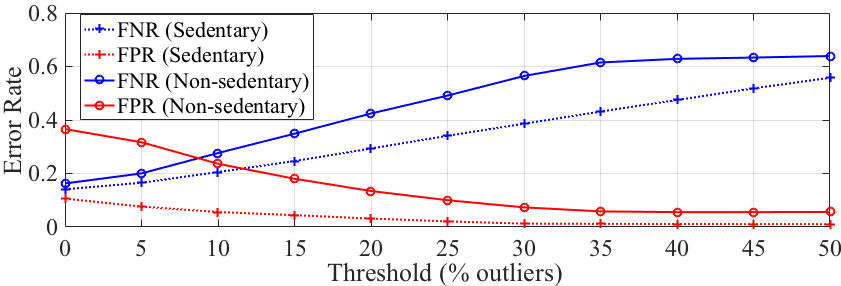}
  \caption{FPR and FNR variations with changing outlier thresholds.}
  \label{fnr_fpr_w5_sed_nonSed_1cls}
\end{figure}

\begin{figure}
\centering
  \includegraphics[width=.95\linewidth]{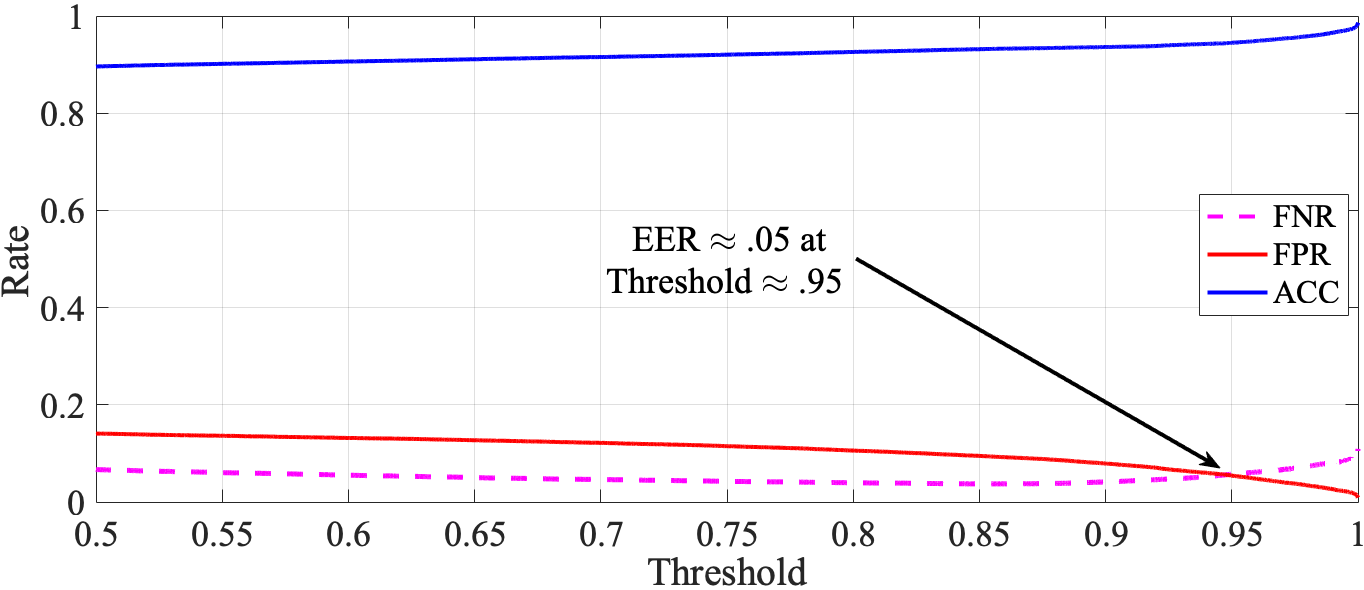}
  \caption{FNR, FPR, and ACC variations with varying probability thresholds for 5 minute windows using the KS-approach with $CM$.}
  \label{w5min_nonSed}
\end{figure}

\ignore{-------------------------------------
\begin{figure}
\centering
\begin{subfigure}{.45\textwidth}
  \centering
  \includegraphics[width=.95\linewidth]{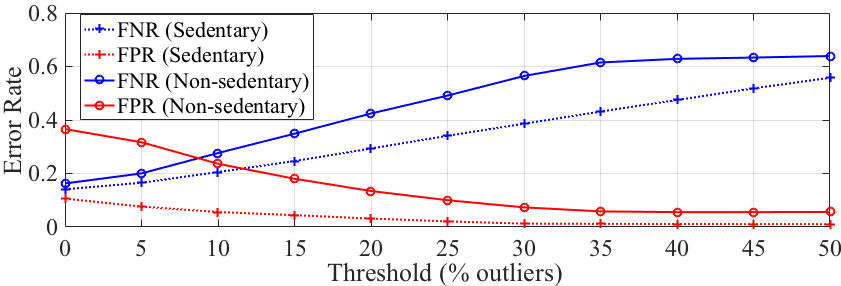}
  \caption{Effect of outlier threshold}
  \label{fnr_fpr_w5_sed_nonSed_1cls}
\end{subfigure}%
\newline
\newline
\begin{subfigure}{.45\textwidth}
  \centering
  \includegraphics[width=.95\linewidth]{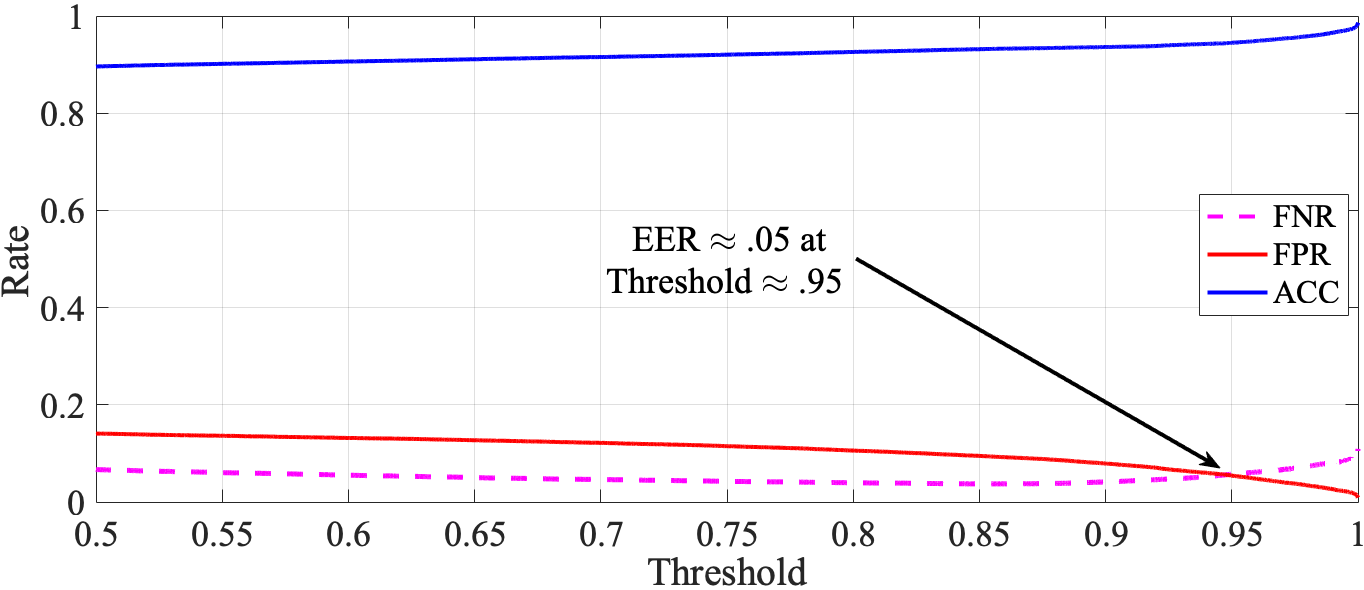}
  \caption{Non-sedentary}
  \label{w5min_nonSed}
\end{subfigure}%
\caption{(a) FPR and FNR variations with changing outlier thresholds. (b) FNR, FPR, and ACC variations with varying probability thresholds for 5 minute windows using the KS-approach with $CM$.}
\label{outlier_th_fnr_fpr_acc_w5min}
\end{figure}
-------------------------------------------}

\subsection{Binary versus Unary Classification}\label{b_vs_u_models}

%\textcolor{blue}{
The unary classifiers in Table~\ref{table:acc_summ_feat_sets} are built without considering any outlier (i.e., 0\% outlier). We observe that the unary models achieve performance close to binary classification models for the KS- and SD-based approaches during sedentary periods. 
However, for non-sedentary periods, the difference between the two classification models is relatively high compared to sedentary periods. 
This is probably due to the variations of window counts across the three non-sedentary activity levels of individuals.
%\textcolor{mycolor}{
In Figure~\ref{fnr_fpr_w5_sed_nonSed_1cls} 
we observe that FPR goes below .05 with $\approx$10\% and $\approx$30\% outliers during sedentary and non-sedentary period models, respectively. 
Therefore, these could be used as appropriate outlier thresholds while designing unary models. 
%}

\subsection{Error Analysis}\label{errAnalysis}
In this section, we take into account the confidence level of a prediction in terms of a posterior probability, which indicates the likelihood of the prediction coming from a particular class. 
In Figure~\ref{w5min_nonSed} we observe that with the increase of the probability threshold, FPR starts dropping sharply, while FNR and ACC increase steadily. After .9 probability all three measures change sharply. At a probability threshold of $\approx$.95, we obtain an EER of $\approx$.05. When FPR drops to .02, FNR increases to $\approx$.1. At that point ACC increases above .97.
For sedentary periods we observe similar patterns. 
This way a trade-off can be made between FPR and FNR to balance security (in terms of FPR) and usability (in terms of FNR) of an implicit authentication system for wearable device users depending on the application scenario and user preference.

\section{Discussion}
\label{discussion}

To our best knowledge, our work is the first to use three different types of less informative coarse-grained processed biometric data (i.e., behavioral, physiological, and hybrid) to accurately authenticate the wearable-users implicitly during both sedentary and non-sedentary periods. 
Our detailed analysis shows the effectiveness and importance of different biometrics and feature selection approaches.

Our activity-level-based models are applicable to any unknown activity type since all types of activities belong to one of the four major activity levels used in our modeling. 
%\textcolor{mycolor}{
Homogeneity of subjects and similarity of their biometrics might have negatively impacted our results. 
Degradation of performance during {\em non-sedentary} periods indicates the need for better modeling, e.g., separate models for different activity levels, which our current dataset does not support due to the lack of sufficient numbers of highly active samples.
%}

%\section*{Acknowledgements}
%The research reported in this paper was supported by the National Heart, Lung, and Blood Institute (NHLBI) of the National Institutes of Health (NIH) under award number R01HL117757.
%The content is solely the responsibility of the authors and does not necessarily represent the official views of the National Institutes of Health.

\ifCLASSOPTIONcaptionsoff
  \newpage
\fi

\bibliographystyle{IEEEtran}
%%\bibliography{reference}
%\bibliography{reference_short}

\bibliography{reference_short}{}

\vfill

\end{document}